\documentclass[prb,aps,twocolumn,superscriptaddress,floatfix,showpacs,10pt]{revtex4-1}
\usepackage{graphicx,amsfonts,amssymb,amsmath,hyperref,dsfont}

\newif\ifhyper
\hypertrue
\ifhyper
\hypersetup{
   citecolor = {green},
   colorlinks = {true}, 
   urlcolor = {blue} 
}
\fi

\newcommand{\beq}{\begin{equation}}
\newcommand{\eeq}{\end{equation}}
\newcommand{\beqa}{\begin{eqnarray}}
\newcommand{\eeqa}{\end{eqnarray}}

\def\Longarrow{\protect\@lra}
\def\@lra{\relbar\joinrel\relbar\joinrel\relbar\joinrel%
          \relbar\joinrel\rightarrow}

\def\be{\begin{equation}}       \def\ee{\end{equation}}
\def\bea{\begin{eqnarray}}      \def\eea{\end{eqnarray}}
\def\bes{\begin{subequations}}  \def\ees{\end{subequations}}

\begin{document}

\title{Quantum phase transitions to topological Haldane phases in spin-one chains studied by linked-cluster expansions}

\author{P. Adelhardt}
\email{patrick.adelhardt@fau.de}
\affiliation{Institute for Theoretical Physics, FAU Erlangen-N\"urnberg, Germany}

\author{J. Gritsch}
\email{julian.gritsch@fau.de}
\affiliation{Institute for Theoretical Physics, FAU Erlangen-N\"urnberg, Germany}

\author{M. Hille}
\affiliation{Lehrstuhl f\"ur Theoretische Physik 1, TU Dortmund, Germany}

\author{D.~A.~Reiss}
\email{david.reiss@fau.de}
\affiliation{Institute for Theoretical Physics, FAU Erlangen-N\"urnberg, Germany}

\author{K.~P.~Schmidt}
\email{kai.phillip.schmidt@fau.de}
\affiliation{Institute for Theoretical Physics, FAU Erlangen-N\"urnberg, Germany}

\date{\rm\today}

\begin{abstract}
 We use linked-cluster expansions to analyze the quantum phase transitions between symmetry unbroken trivial and topological Haldane phases in two different spin-one chains. The first model is the spin-one Heisenberg chain in the presence of a single-ion anisotropy while the second one is the dimerized spin-one Heisenberg chain. For both models we determine the ground-state energy and the one-particle gap inside the non-topological phase as a high-order series using perturbative continuous unitary transformations. Extrapolations of the gap series are applied to locate the quantum critical point and to extract the associated critical exponent. We find that this approach works unsatisfactory for the anisotropic chain, since the quality of the extrapolation appears insufficient due to the large correlation length exponent. In contrast, extrapolation schemes display very good convergence for the gap closing in the case of the dimerized spin-one Heisenberg chain.
\end{abstract}
 

\maketitle

%
%
\section{Introduction}
%
%
Topologically ordered quantum phases have attracted an enormous interest in recent years due to their fascinating physical properties. Such phases display long-range quantum entanglement in the ground state and support exotic excitations with fractional quantum numbers as well as, in two dimensions, featuring unconventional particle statistics different from conventional fermions or bosons. The latter excitations called anyons \cite{leinaas77,wilczek82} are at the heart of topological quantum computation \cite{kitaev03,freedman03,nayak08}. Quantum phases with such topological order are robust against small quantum fluctuations. But strong enough perturbations destroy the topological order via a quantum phase transition to a different ground state, which is usually not topologically ordered. Since topological order cannot be characterized by local order parameters, these quantum phase transitions cannot be described by Landau's paradigm of phase transitions. Therefore, it is interesting and important to investigate such quantum critical behavior.    

One relevant arena to explore topological quantum phase transitions are interacting quantum spin systems, which are known to realize topological order and associated phase transitions in a large variety of microscopic models and dimensions. This includes three-dimensional quantum spin-ice models displaying a quantum phase transition out of a topological Coulomb phase \cite{savary17,roechner16}, two-dimensional toric codes, Kitaev and string-net models showing a plethora of phase transitions in the presence of additional perturbations \cite{levin05,trebst07,hamma08,vidal09_1,vidal09_2,gils09,tupitsyn10,dusuel11,wu12,schmi13,schulz13,schulz14,schulz16} as well as one-dimensional quantum spin chain models. 

The most prominent one-dimensional spin system displaying topological order is the antiferromagnetic spin-one chain \cite{haldane83a,haldane83b,affleck87,pollmann12}. Its ground state possesses long-range string order \cite{nijs89,kennedy92} and its elementary excitations above the topological singlet ground state are gapped consistent with Haldane's conjecture \cite{haldane83a,haldane83b,affleck87}. A quantum phase transition out of such a Haldane phase can be induced by additional interactions, e.g.~, a single-ion anisotropy \cite{botet83,schulz86} or a dimerization giving rise to the dimerized spin-one Heisenberg chain. It is known that the quantum phase transition between the gapped topological Haldane phases and, in both models, topologically trivial gapped phases is continuous, belonging to the Gaussian universality class \cite{kitazawa96,chen03,degli03,tzeng08,ueda08,hu11}.

In this work we study this topological phase transition of the antiferromagnetic spin-one Heisenberg chain in the presence of single-ion anisotropy or dimerization. Our main interest is whether one can understand the continuous phase transition by the closing of the one-particle gap of the trivial phases, i.e.~, from the limit of infinitely strong dimerization or single-ion anisotropy where the ground state is given by unentangled product states. To this end we set up high-order linked-cluster expansions for the one-particle excitations inside the trivial phases using the method of perturbative continuous unitary transformations (pCUTs) \cite{knetter00,knetter03}. Extrapolations of the one-particle gap allow to locate the quantum critical points and to extract the associated critical exponents. We find that the extrapolation of the one-particle gap yields unsatisfying results for the spin-one Heisenberg chain in the presence of a single-ion anisotropy. This likely originates from the fact that the correlation length exponent is larger than one \cite{hu11} so that the one-particle gap closing is very flat.  In contrast, our results for the dimerized spin-one Heisenberg chain compare well with the other numerical resulst of the literature, which we discuss at the end of this work. Consequently, our results provide an confirmation of previous results, obtained by complementary series expansion methods.

The paper is organized as follows. We introduce the microscopic models in Sect.~\ref{sec::model} and we explain all the technical aspects in Sect.~\ref{sec::pCUT}. This includes the pCUT method as well as the applied extrapolation schemes. In Sect.~\ref{sec::results}, we present and discuss our results. Finally, we draw conclusions in Sect.~\ref{sec::conclusion}.  
%
%
\section{Models}
\label{sec::model}
\begin{figure}[t]
	\centering
		\includegraphics[width=\columnwidth]{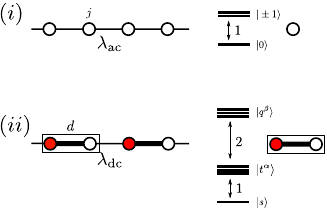}
	\caption{(i) The left part illustrates the Heisenberg chain in the presence of a single-ion anisotropy. Circles embody spin-one degrees of freedom and black lines represent Heisenberg interactions with strength $\lambda_{\rm ac}$. The right part shows the energy spectrum of an isolated spin subject to a single-ion anisotropy. The three eigenstates are denoted by $|\alpha\rangle$ with $\alpha\in\{0,\pm 1\}$ labelling the quantum number of $S^{z}$. (ii) The left part illustrates the dimerized Heisenberg chain. The black box shows the dimer $d$ consisting of a left (red circles) and a right (white circles) spin one. Black lines represent inter-dimer Heisenberg interactions with strength $\lambda_{\rm dc}$. On the right the energy spectrum of an isolated spin-one Heisenberg dimer is sketched. The lowest-energy eigenstate is the singlet $|s\rangle$. Excitations are triplet states $|t^\alpha\rangle$ with $\alpha\in\{ 0,\pm 1\}$ and quintuplet states $|q^\beta\rangle$ with $\beta\in\{ 0,\pm 1,\pm 2\}$.}
	\label{fig:models}
\end{figure}

%
%
We consider two types of antiferromagnetic spin-one chains: i) a Heisenberg chain in the presence of a single-ion anisotropy (AC) and ii) a dimerized Heisenberg chain (DC). Both models are illustrated in Fig.~\ref{fig:models}.

%

%
\subsection{Single-ion anisotropy}
\label{ssec::model_ac}
%
The Hamiltonian of the Heisenberg chain in the presence of a single-ion anisotropy reads 
\begin{equation}
\mathcal{H}^{\rm (ac)} =  \sum_{j}  \left( S_{j}^{\rm z}\right)^2  + \lambda_{\rm ac}\sum_{j}  \boldsymbol{S}_{j}\cdot \boldsymbol{S}_{j+1}\quad .
 \label{eq:ham_single_ion}
\end{equation}
The sums run over all sites $j$ of a one-dimensional chain, $\lambda_{\rm ac}\in[0,\infty)$, and $\boldsymbol{S}_{j}=(S_{j}^{\rm x},S_{j}^{\rm y},S_{j}^{\rm z})$ represents the spin-one operator on site $j$. For large $\lambda_{\rm ac}$, this model realizes a topological Haldane phase characterized by a non-local string order parameter and gapped elementary excitations. In contrast, for small $\lambda_{\rm ac}$, one finds a ground state which is adiabatically connected to the case $\lambda_{\rm ac}=0$ of decoupled spin-ones. The three spin eigenstates in the presence of a single-ion anisotropy are denoted by $|\alpha\rangle$ with $\alpha\in\{0,\pm 1\}$ labelling the quantum number of $S^{z}$. The eigenenergy of state $|\alpha\rangle $ is $\alpha^2$ as sketched on the right side of Fig.~\ref{fig:models}(i). For $\lambda_{\rm ac}=0$, the ground state is the unentangled product state $\prod_j |0 \rangle_j$ and elementary excitations are local spin flips to the states $|\pm 1\rangle$ of single spins costing an energy $\Delta^{\rm (ac)}=1$. Therefore, for small $\lambda_{\rm ac}$, the phase is featureless without symmetry breaking and has gapped excitations. Consequently, there must occur a quantum phase transition between the two gapped phases when the parameter $\lambda_{\rm ac}$ is varied. This Gaussian transition has been located accurately by density matrix renormalization group calculations to be at $\lambda_{\rm ac}^{\rm c}\approx 1.03258$ and the gap in both quantum phases closes with a critical exponent $z\nu\approx 1.472$ \cite{hu11}. Here $z$ is the dynamical and $\nu$ the correlation length exponent. 

%
\subsection{Dimerized chain}
\label{ssec::model_dc}
%

The Hamiltonian of the dimerized Heisenberg chain is given by
\begin{equation}
\mathcal{H}^{\rm (dc)} =  \sum_{d}  \boldsymbol{S}_{d, L}\cdot \boldsymbol{S}_{d, R} + \lambda_{\rm dc}\sum_{d} \boldsymbol{S}_{d, R}\cdot \boldsymbol{S}_{d+1, L}\quad ,
 \label{eq:ham_dim_chain}
\end{equation}
%
where $\boldsymbol{S}_{d, L}$ ($\boldsymbol{S}_{d, R}$) denotes a spin-one operator on the left (right) site of dimer $d$ (see also Fig.~\ref{fig:models}(ii)). The parameter $\lambda_{\rm dc}\in [0,1]$ varies from the isolated dimer limit $\lambda_{\rm dc}=0$ to the isotropic Heisenberg chain $\lambda_{\rm dc}=1$. The ground state for $\lambda_{\rm dc}=0$ is the unentangled product state $\prod_d |s\rangle_d$ where $|s\rangle_d$ is the singlet eigenstate of the isolated dimer $d$. Elementary excitations are local triplets $|t^\alpha\rangle$ with $\alpha\in\{ 0,\pm 1\}$ and quintuplets $|q^\beta\rangle$ with $\beta\in\{ 0,\pm 1,\pm 2\}$ as illustrated on the right side of Fig.~\ref{fig:models}(ii). In contrast, the isotropic Heisenberg chain at $\lambda_{\rm dc}=1$ is known to be in a topologically ordered gapped Haldane phase like the AC for large $\lambda_{\rm ac}$ discussed above. Thus a Gaussian quantum phase transition must occur as a function of $\lambda_{\rm dc}$ between the two gapped phases, which is known to take place at $\lambda_{\rm dc}^{\rm c}\approx 0.6$ \cite{singh88,kato94,yamamoto95,yamamoto95b,totsuka95,pati96,chandra06,liu14,su16}. The associated gap closing critical exponent is $z\nu=1$ \cite{haldane83a,haldane83b,botet83,schulz86,affleck87,kato94,yamamoto95,yamamoto95b,su16}. 

%
%
\section{pCUT}
\label{sec::pCUT}
%
%
In this section we provide the relevant technical aspects of pCUT applied to both spin-one chains. We start by sketching the underlying method of the expansion; for details the reader may consult Refs.~\onlinecite{knetter00,knetter03,coester15}. The expansion's reference point is $\lambda_{\kappa}=0$ with $\kappa\in\{{\rm ac,dc}\}$. Here the ground state is given by a product state in both models. The spin-one Heisenberg chain in the presence of single-ion anisotropy is in the state where each spin on site $j$ is in the state $|0\rangle_j$, and the elementary excitations are local excitations $|\pm 1\rangle_j$ having $S^{\rm z}_j=\pm 1$ and an excitation energy $\Delta^{\rm (ac)}=1$ (see Fig.~\ref{fig:models}(i)). For the DC, isolated dimers are in the singlet state $|s\rangle_d$, and elementary excitations are local triplets with total spin 1 and with excitation energy $\Delta^{\rm (dc)}=1$ as well as local quintuplet excitations with total spin 2 and excitation energy 3 as illustrated in Fig.~\ref{fig:models}(ii).  

After a global energy shift, we can rewrite both models in the form
\begin{eqnarray}
\label{h_pert}
\mathcal{H^{(\kappa )}}&=&\mathcal{H}^{(\kappa )}_0+\lambda_{\kappa}\, \hat{V}^{(\kappa )} \quad ,
\end{eqnarray}
where $\kappa\in\{{\rm ac,dc}\}$ and $\mathcal{H}_0^{(\kappa )}$ is a counting operator of elementary energy quanta. The number of energy quanta is equal to the number of local excitations $|\pm 1\rangle$ with $S^{\rm z}=\pm 1$ for the AC. In contrast, for the DC, the number of energy quanta is given by the number of triplet excitations $|t^{\alpha}\rangle$ with $\alpha\in\{0,\pm 1\}$ plus three times the number of quintuplet excitations $|q^{\beta}\rangle$ with $\beta\in\{0,\pm 1,\pm 2\}$, since the eigenenergy of the states  $|q^{\beta}\rangle$ is 3.  

The perturbations can be written as
\begin{align}
\hat{V}^{(\rm ac )}=\hat{T}^{(\rm ac )}_{-2}+\hat{T}^{(\rm ac )}_0+\hat{T}^{(\rm ac) }_{2} 
\end{align}
for the AC and 
\begin{align}
\hat{V}^{\rm (dc)}=\sum_{m=-4}^4 \hat{T}^{(\rm dc )}_{m} 
\end{align}
for the DC where $\hat{T}_m^{(\kappa)}$ changes the total number of energy quanta by $m$. 

Each operator $\hat{T}_m^{(\kappa)}$ is a sum over local operators connecting two nearest-neighbor supersites on the chain, where a supersite corresponds to a single spin for the AC and a dimer for the DC. One can therefore write
\begin{align}
\hat{T}^{(\kappa)}_m=\sum_l \hat{\tau}^{(\kappa)}_{m,l}\quad ,\label{tausumme}
\end{align}
with $\hat{\tau}_{m,l}^{(\kappa)}$ affecting the two supersites connected by the link $l$ on the chain of supersites.

The pCUT method \cite{knetter00,knetter03,coester15} maps the original Hamiltonian to an effective quasiparticle-conserving Hamiltonian of the form
\begin{eqnarray}
\mathcal{H}^{(\kappa)}_\text{eff} =\mathcal{H}^{(\kappa)}_0+\sum_{n=1}^{\infty}\lambda_{\kappa}^n \hspace*{-2mm}
\sum_{{\rm dim}(\underline{m})=n \atop  \,M(\underline{m})=0} \hspace*{-2mm}
C(\underline{m})\,\hat{T}_{m_1}^{(\kappa)}\dots \hat{T}^{(\kappa)}_{m_n},\label{H_eff}
\end{eqnarray}
where $n$ reflects the perturbative order. The second sum runs over all possible vectors $\underline{m}\!\equiv\!(m_1, \ldots, m_n)$ with $m_i\in\{\pm 2, 0\}$ ($m_i\in\{\pm 4,\pm 3, \pm 2, \pm 1, 0\}$) for the AC (DC) and dimension ${\rm dim}(\underline{m})=n$. Each term of this sum is weighted by the rational coefficient \mbox{$C(\underline{m})\in \mathbb{Q}$} which has been calculated model-independently up to high orders.\cite{knetter00} The additional restriction $M(\underline{m})\equiv\sum_i m_i=0$ reflects the quasiparticle-number conservation of the effective Hamiltonian, i.e.~, the resulting Hamiltonian is block-diagonal in the number of energy quanta, \mbox{$[\mathcal{H}^{(\kappa)}_\text{eff},\mathcal{H}^{(\kappa)}_0]=0$.} Each quasiparticle-number block can be investigated separately which represents a major simplification of the complicated many-body problem.

The operator products $\hat{T}_{m_1}\dots \hat{T}_{m_n}$ appearing in order $n$ can be interpreted as virtual fluctuations of ``length'' $l\leq n$ leading to dressed quasiparticles. According to the linked-cluster theorem, only linked fluctuations can have an overall contribution to the effective Hamiltonian $\mathcal{H}_\text{eff}$. Hence, the properties of interest can be calculated in the thermodynamic limit by applying the effective Hamiltonian to finite chain segments. 

In practice, we calculated high-order series expansions for the zero- and one-quasiparticle sector for both models.  Note that the computations for the DC are more demanding than the ones for the AC. Reasons are the larger local Hilbert space of a dimer compared to that of a single spin and the larger number of operators $\hat{T}_m$ resulting in more operator sequences in the effective Hamiltonian \eqref{H_eff}. As a consequence, we reach lower perturbation orders for the DC compared to the AC. Additionally, the coefficients of the series are obtained as exact fractions for the AC while we had to calculate with float number for the DC. The zero-quasiparticle sector yields directly the ground-state energy per supersite which we calculated up to order $14$ ($8$) for the AC (DC). Similar calculations in the one-quasiparticle sector result in the one-particle hopping amplitudes which we determined up to order $15$ ($8$) for the AC (DC). A Fourier transformation diagonalizes the one-particle hopping Hamiltonian for both spin-one chains. This yields the one-particle dispersion $\omega^{(\kappa)}(k)$ and the one-particle gap $\Delta^{(\kappa)}\equiv \omega^{\rm (\kappa)}(k=0)$. 

%
\subsection{Extrapolation}
\label{ssec::extrapol}
%
In order to detect second-order quantum phase transitions, we use Pad\'e and DlogPad\'e techniques to extrapolate the one-particle gap $\Delta^{\rm (\kappa)}$ \cite{guttmann89}. To this end, various extrapolants $\left[L,M\right]$ are constructed, where $L$ denotes the order of the numerator and $M$ the order of the denominator. 

A standard extrapolation scheme is the Pad\'e extrapolation which is defined by 
\begin{align}
\label{pade}
{\rm P}[L,M]_{\Delta}\equiv\frac{P_L(\lambda)}{Q_M(\lambda)}=\frac{p_0+p_1\lambda+\dots + p_L \lambda^L}{q_0+q_1\lambda+\dots +q_M \lambda^M}\quad,
\end{align}
for the one-particle gap $\Delta$ and with $p_i,q_i \in \mathbb{R}$ and \mbox{$q_0=1$.} The latter can be achieved by reducing the fraction. The real coefficients are fixed by the condition that the Taylor expansion of ${\rm P}[L,M]_{\Delta}$ about $\lambda=0$ up to order \mbox{$r\equiv L+M$} with \mbox{$L+M\leq n$} recovers the corresponding Taylor polynomial of order $r$ for the original series of $\Delta$. Here $n$ denotes the maximum perturbative order which has been calculated.

The DlogPad\'e extrapolation is based on the Pad\'{e} extrapolation ${\rm P}[L,M]_{\mathcal{D}}$ of the logarithmic derivative \mbox{$\mathcal{D}(\lambda )\equiv \frac{{\rm d}}{{\rm d}\lambda}\ln \Delta^{\rm (\kappa )}$} of the one-particle gap $\Delta^{\rm (\kappa)}$. Due to the derivative in $\mathcal{D}$ one requires $L+M\leq n-1$. The DlogPad\'e extrapolant ${\rm dP}\left[L,M\right]_{\Delta}$ is then given by
\begin{equation}
 {\rm dP}\left[L,M\right]_{\Delta} \equiv \exp\left(\int_0^\lambda {\rm P}[L,M]_{\mathcal{D}} \,{\rm d}\lambda'\right)\quad .
 \label{eq:dlog2}
\end{equation}
In the case of a physical pole of ${\rm P}[L,M]_{\mathcal{D}}$ at $\lambda^{\rm c}$, which corresponds to the closing of the gap and therefore to the location of the quantum critical point, one is able to determine the dominant power-law behavior $|\lambda-\lambda^{\rm c}|^{z\nu}$ close to $\lambda^{\rm c}$. The critical exponent $z\nu$ is given by the residuum of ${\rm P}[L,M]_{\mathcal{D}}$ at $\lambda=\lambda^{\rm c}$
\begin{equation}
 z\nu = \frac{P_{L}(\lambda )}{\frac{d}{d\lambda }Q_M(\lambda )} {\Bigg|}_{\lambda =\lambda^{\rm c}}\quad ,
 \label{eq:exp}
\end{equation}

where $P_{L}(\lambda)$ ($Q_{M}(\lambda)$) is the numerator (denominator) of ${\rm P}[L,M]_{\mathcal{D}}$. If the exact value of $\lambda^{\rm c}$ is known, one can obtain better estimates of the critical exponent $z\nu$ by defining
\begin{align*}
z\nu^*(\lambda)&\equiv(\lambda^{\rm c}-\lambda)\cdot \mathcal{D}(\lambda )\\
&\approx z\nu+\mathcal{O}(\lambda-\lambda^{\rm c})
\end{align*}
and by then applying standard Pad\'{e} extrapolation on the function $z\nu^*(\lambda)$ 
\begin{align}
{\rm P}[L,M]_{z\nu^*} {\bigg|}_{\lambda =\lambda^{\rm c}}=z\nu \label{biasnue}
\end{align}
evaluated at $\lambda=\lambda^{\rm c}$.

In general, one expects that the quality of the extrapolation increases with the perturbative order. Convergence of a physical quantity is indicated by different extrapolants $\left[L,M\right]$ (Pad\'e or DlogPad\'e extrapolation) and especially different families of extrapolants with $L-M={\rm const}$ yielding the same result. Here and in the following we omit the index $\Delta$ in $\left[L,M\right]_\Delta$ for the sake of brevity. Note that Pad\'e and DlogPad\'e extrapolants can possess so-called spurious poles, i.e.~, poles for $0<\lambda<\lambda^{\rm c}$ or $\lambda\approx \lambda^{\rm c}$, where $\lambda^{\rm c}$ corresponds to the location of a quantum critical point. These spurious poles usually spoil the quality of the extrapolation. Thus associated extrapolants are excluded in the further analysis.     

%
%
\section{Results}
\label{sec::results}
%
%

In this section we present our results and discuss implications for the quantum phase transition between the trivial and the Haldane phases. To this end we focus on the one-particle gap inside the trivial phases of both models, since extrapolations of the ground-state energy did not provide any quantitative indications for the quantum critical behavior. 

%
\subsection{Anisotropic chain}
\label{ssec::ac}
%

As outlined above, we have applied the pCUT method to calculate the ground-state energy per site $\epsilon^{(\rm ac)}_{0}$ as well as the one-particle gap in the form of a high-order series in $\lambda_{\rm ac}$. The explicit expressions are given by
%
\begin{figure}[ht!]
	\centering
		\includegraphics[width=\columnwidth]{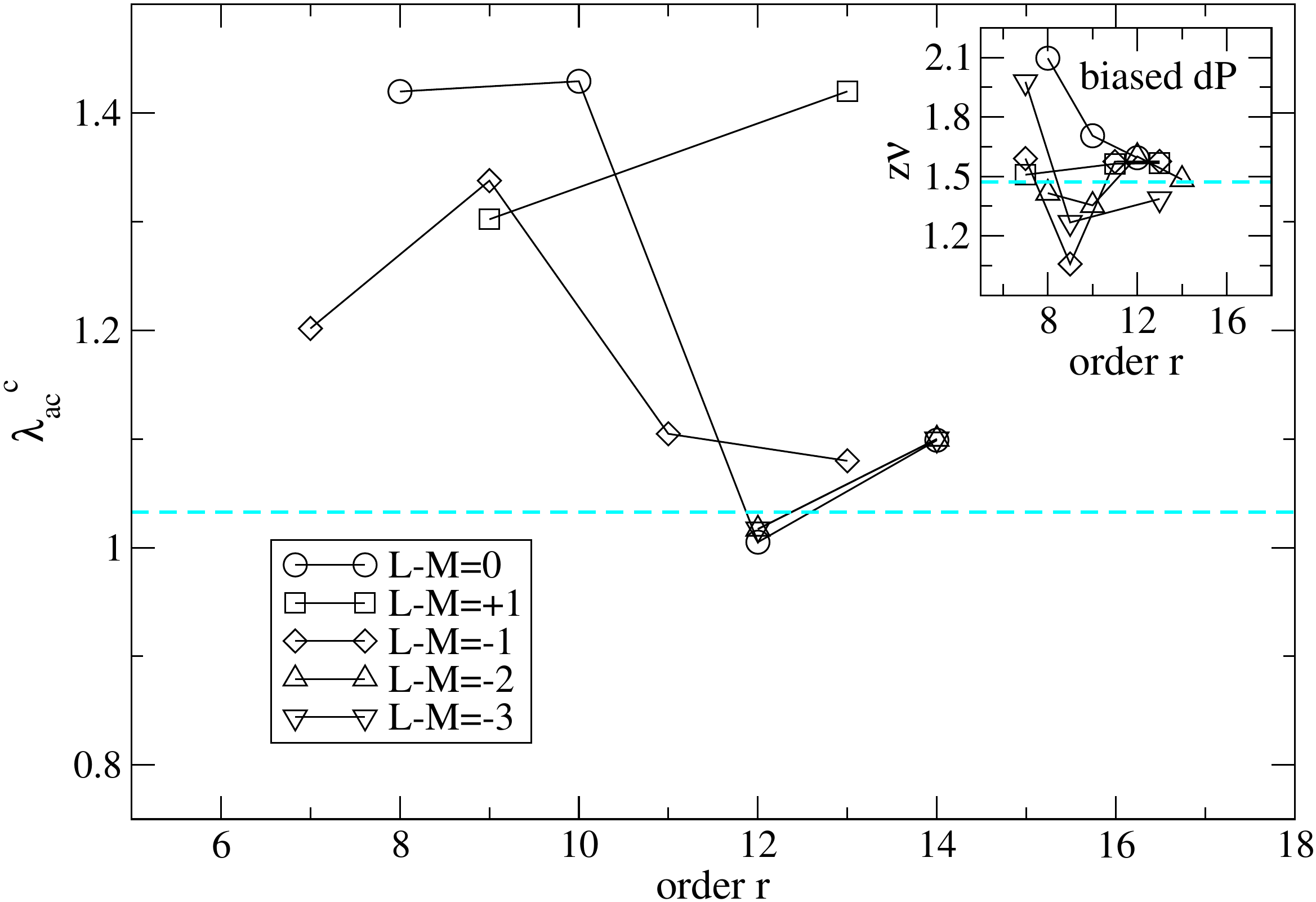}
	\caption{Critical point $\lambda_{\rm ac}^{\rm c}$ as a function of the order $r$ extracted from different families of DlogPad\'{e} extrapolants ${\rm dP}[L,M]$ with $L-M={\rm const}$. The dashed horizontal line indicates $\lambda_{\rm ac}^{\rm c}=1.03258$ from Ref.~\onlinecite{hu11}. {\it Inset}: Critical exponent $z\nu$ extracted from biased DlogPad\'{e} extrapolation using $\lambda_{\rm ac}^{\rm c}=1.03258$ in Eq.~\ref{biasnue}. The dashed horizontal line indicates $z\nu=1.472$ (taken also from Ref.~\onlinecite{hu11}).}
	\label{fig:gap_ac}
\end{figure}
\begin{eqnarray}
        \epsilon^{(\rm ac)}_{0}&=&-\lambda_{\rm ac}^2-\frac{1}{2}\lambda_{\rm ac}^3+\frac{1}{4}\lambda_{\rm ac}^4+\frac{9}{8}\lambda_{\rm ac}^5+\frac{17}{16}\lambda_{\rm ac}^6 -\frac{39}{32}\lambda_{\rm ac}^7  \nonumber\\
                            &&  -\frac{597}{128}\lambda_{\rm ac}^8 -\frac{117}{32}\lambda_{\rm ac}^9 +\frac{52681}{6144}\lambda_{\rm ac}^{10} +\frac{157237}{6144}\lambda_{\rm ac}^{11}   \nonumber\\
                            &&+\frac{11698951}{884736}\lambda_{\rm ac}^{12} -\frac{353210417}{5308416}\lambda_{\rm ac}^{13} \nonumber\\
                            && -\frac{20232615041}{127401984}\lambda_{\rm ac}^{14}
\label{eq:AC_0QP}
\end{eqnarray}
%
and
\begin{eqnarray}
        \Delta^{(\rm ac)} &=&1-2\lambda_{\rm ac} +\lambda_{\rm ac}^2+\frac{1}{2}\lambda_{\rm ac}^3-\frac{3}{4}\lambda_{\rm ac}^4-\frac{1}{4}\lambda_{\rm ac}^5+\frac{3}{32}\lambda_{\rm ac}^6   \nonumber\\
                            &&  +\frac{99}{32}\lambda_{\rm ac}^7-\frac{53}{128}\lambda_{\rm ac}^8 -\frac{14367}{2048}\lambda_{\rm ac}^9 -\frac{11647}{768}\lambda_{\rm ac}^{10}   \nonumber\\
                            &&+\frac{4659605}{294912}\lambda_{\rm ac}^{11} +\frac{257499161}{3538944}\lambda_{\rm ac}^{12} \nonumber\\
                            &&+\frac{3026827735}{42467328}\lambda_{\rm ac}^{13} -\frac{3056050607}{14155776}\lambda_{\rm ac}^{14}\nonumber\\
                            &&-\frac{982259794445}{1528823808}\lambda_{\rm ac}^{15}\quad . 
\label{eq:AC_1QP}
\end{eqnarray}
%

Next we analyze the one-particle gap to extract the quantum critical properties of the AC by applying DlogPad\'{e} extrapolation. The resulting quantum critical points $\lambda_{\rm ac}^{\rm c}$ are shown for various families of extrapolants ${\rm dP}[L,M]$ with $L-M={\rm const}$ as a function of the total order $r=L+M$ in Fig.~\ref{fig:gap_ac}.

Obviously, the quality of the extrapolation is not very convincing, since the locations of the critical point scatter between $\lambda_{\rm ac}^{\rm c}\approx 1$ and $\lambda_{\rm ac}^{\rm c}\approx 1.5$. We attribute this to the gap expected to close with an exponent $z\nu=1.472>1$ \cite{hu11}. As a consequence, the gap closing is very flat. It is reasonable to interpret the unsatisfactory extrapolation as being caused by the fact that it is hard for the DlogPad\'{e} extrapolation to locate a very flat gap closing precisely. Indeed, if one studies the associated critical exponents for the extrapolants shown in Fig.~\ref{fig:gap_ac}, one finds values for $z\nu$ in the large interval $1$ to $5$ (not shown). We therefore biased the DlogPad\'{e} extrapolants to the critical value $\lambda_{\rm ac}^{\rm c}=1.03258$ \cite{hu11} and extracted the critical exponent using Eq.~\eqref{biasnue}. These results are displayed in the inset of Fig.~\ref{fig:gap_ac}. We first note that a larger number of DlogPad\'{e} extrapolants do not show spurious poles when biasing with $\lambda_{\rm ac}^{\rm c}=1.03258$. Furthermore, the different families of extrapolants seem to converge to the critical exponent $z\nu\approx 1.5$ consistent with the expected value $z\nu=1.472>1$ \cite{hu11}. Still, the overall quality of the extrapolation is not sufficient to gain any quantitative insights into the quantum critical behavior of the AC.   

%
\subsection{Dimerized chain}
\label{ssec::dc}
%

\begin{figure}[t]
	\centering
		\includegraphics[width=\columnwidth]{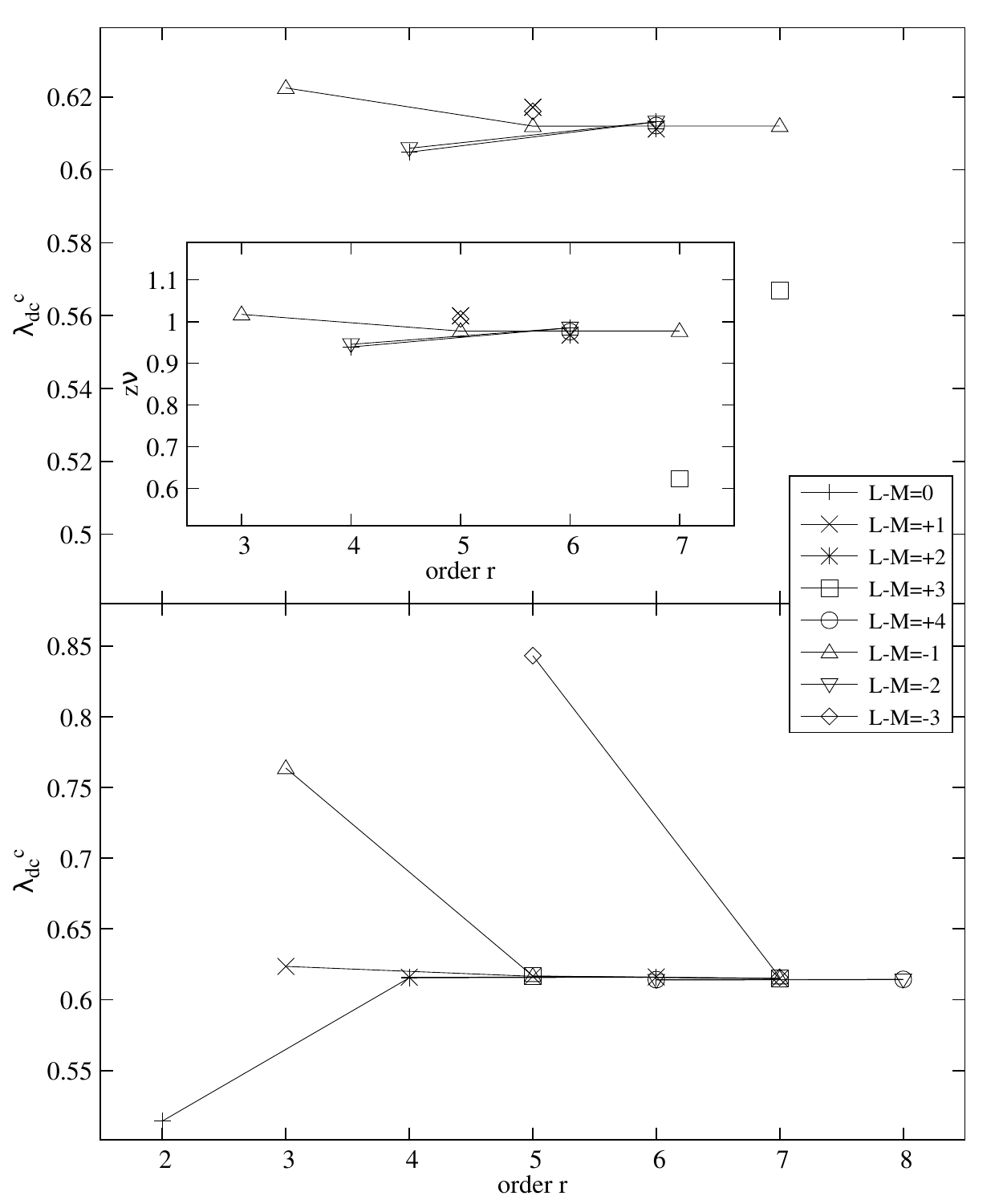}
	\caption{The critical point $\lambda_{\rm dc}^{\rm c}$ displayed as a function of the order $r$ in the upper (lower) panel using DlogPad\'e (Pad\'{e}) extrapolation. Different symbols correspond to different families of extrapolants with $L-M={\rm const}$. {\it Inset}: Critical exponent $z\nu$ extracted from DlogPad\'{e} extrapolation.}
	\label{fig:extrapol}
\end{figure}

Next we look at the DC. The series expansion for the ground-state energy reads
\begin{eqnarray}
        \epsilon^{(\rm dc)}_{0} &=&- 0.6666666667 \;\lambda_{\rm dc}^2 - 0.1666666666 \;\lambda_{\rm dc}^3 \nonumber\\
        &&+ 0.0092592592 \;\lambda_{\rm dc}^4 - 0.0413580247 \;\lambda_{\rm dc}^5 \nonumber\\
        &&- 0.0304486740 \;\lambda_{\rm dc}^6 + 0.0054201910 \;\lambda_{\rm dc}^7 \nonumber\\
        &&- 0.0237290153 \;\lambda_{\rm dc}^8
\label{eq:DC_0QP}
\end{eqnarray}
%
and the one-particle gap is given by 
\begin{eqnarray}
 \Delta^{\rm (dc)} &=& 1.000000000 - 1.3333333333\;\lambda_{\rm dc}  \nonumber\\
         && - 0.8148148149\;\lambda_{\rm dc}^2 + 1.1481481483\;\lambda_{\rm dc}^3 \nonumber\\
	 &&  - 1.9746513492\;\lambda_{\rm dc}^4 + 3.4647814745\;\lambda_{\rm dc}^5 \nonumber\\
         && - 6.4440281143\;\lambda_{\rm dc}^6 +  12.3431841724\;\lambda_{\rm dc}^7 \nonumber\\
         && - 27.6026443473\;\lambda_{\rm dc}^8\quad .
\label{eq:DC_1QP}
\end{eqnarray}
%
For this model one expects that the gap closes around $\lambda_{\rm dc}^{\rm c}\approx 0.6$ with a critical exponent $z\nu=1$. We have applied Pad\'e and DlogPad\'e extrapolation to the one-particle gap in order to study the quantum phase transition of the DC. The results are shown in Fig.~\ref{fig:extrapol}.
%
In contrast to the AC, we find a much better convergence for the DC in both extrapolation schemes. The DlogPad\'e extrapolants converge to a value $\lambda_{\rm dc}^{\rm c}\approx 0.61$. An exception is extrapolant ${\rm dP}[5,2]$ (square symbol for $r=7$) which is the single member of the rather non-diagonal family ${\rm dP}[L,M]$ with $L-M=3$ and might therefore be ignored. The good convergence of all the other extrapolants is also reflected in the associated critical exponents which are shown in the inset of Fig.~\ref{fig:extrapol}. Averaging over the extrapolants with the highest reliable order $r=6$ gives the critical exponent $z\nu=0.98\pm 0.01$ consistent with the expected behavior. The single extrapolant with $r=7$ is less reliable, because one  cannot compare it to other extrapolants of the same order. 

%
\begin{figure}[ht]
	\centering
		\includegraphics[width=\columnwidth]{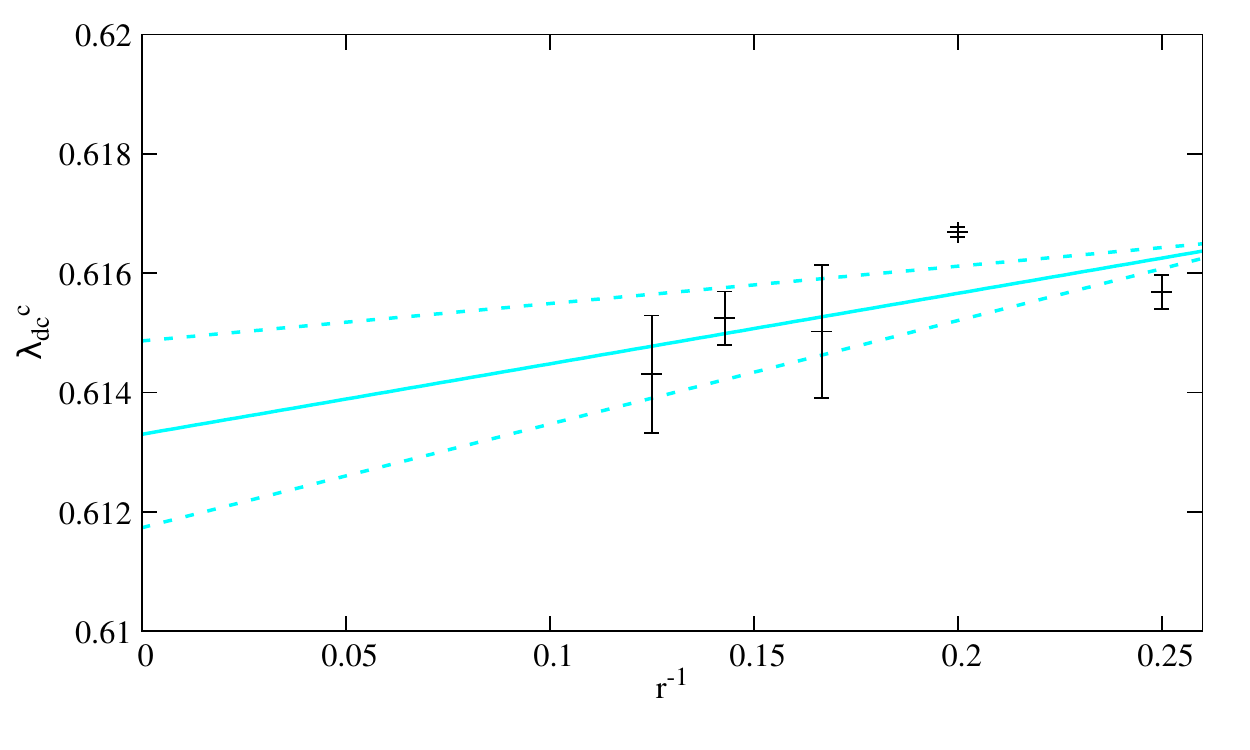}
	\caption{Linear scaling of the critical point $\lambda_{\rm dc}^{\rm c}$ as a function of $1/r$. Data points are the average of Pad\'e extrapolants of the same order shown in Fig.~\ref{fig:extrapol}. Error margins correspond to their variance. Solid and dashed lines are linear scalings through the data points and their variances, respectively.}
	\label{fig:scaling}
\end{figure}

The linear closing of the gap can also be well described by a standard Pad\'e extrapolation. We find a good convergence as can be seen from the lower panel of Fig.~\ref{fig:extrapol}. All families converge to a critical value close to 0.61. In order to get a more quantitative estimate of the quantum critical point, we calculated average values of the Pad\'e extrapolants of the same order $r$ and scaled the averages to the infinite-order limit $r\rightarrow \infty$. The results of this scaling are shown in Fig.~\ref{fig:scaling} as a function of $1/r$. For this we have omitted values from too small orders $r<4$ as well as families' small-order outliers deviating strongly from high-order members of the same family. A linear scaling of the averaged values and their variances gives $\lambda_{\rm dc}^{\rm c}=0.6133\pm 0.0016$. Note that the error bar reflects the rather small variance of different Pad\'e extrapolants and does not correspond to an error margin in the sense of measurement uncertainties or numerical errors.

Let us compare our findings to the results for the DC in the literature. The critical point at \mbox{$\lambda_{\rm dc}^{\rm c}\approx 0.61$} and the critical exponent $z\nu \approx 1$ is consistent with other investigations. High-order series expansions for physical quantities other than the gap amounts to $\lambda_{\rm dc}^{\rm c}=0.6 \pm 0.04$ \cite{singh88}. Density matrix renormalization group calculations yield $\lambda_{\rm dc}^{\rm c}=0.6\pm 0.0128$ by analyzing the gap closing \cite{kato94,pati96}. Similarly, quantum Monte Carlo simulations obtain $\lambda_{\rm dc}^{\rm c}\approx 0.6$ via the one-particle gap, extracting also the effective light velocity $v\approx 2.46\pm 0.01$ as well as the central charge $c\approx 1.02\pm 0.09$ at the quantum critical point \cite{yamamoto95,yamamoto95b}. Interestingly, other works point towards a critical value slightly below $0.6$ \cite{totsuka95,chandra06,liu14,su16}. Exact diagonalization gives $\lambda_{\rm dc}^{\rm c}=0.595 \pm 0.01$ \cite{totsuka95} when performing finite-size scaling of the Binder parameter with the non-local string order parameter. The infinite time-evolving block decimation (iTEBD) method yields $\lambda_{\rm dc}^{\rm c}= 0.587 \pm 0.002$ by analyzing the singular behavior of the bipartite entanglement in the ground state \cite{liu14,su16}. Altogether, it seems that estimators of ground-state properties like the bipartite entanglement or the non-local string order parameter tend to smaller values $\lambda_{\rm dc}^{\rm c}\approx 0.59$ compared to other investigations (including ours) which locate the quantum critical point via the closing of the excitation gap. 
%
%
\section{Conclusions}
\label{sec::conclusion}
%
%
We calculated high-order series expansions for the ground-state energy per site and the one-particle gap for two different spin-one chains using the method of perturbative continuous unitary transformations. In both cases the expansion is performed inside a trivial phase where the unperturbed reference state is a product state of isolated spins (AC) or isolated dimers (DC). Computationally, the expansion for the DC is more involved due to the larger local Hilbert space of a single dimer compared to that of a single spin-one for the AC. As a consequence, substantially higher orders are reached for the AC. 

Both spin-one chains display a second-order quantum phase transition between two symmetry unbroken gapped ground states. The systems are either in a trivial phase or in a topological Haldane phase with unconventional non-local string order parameter. Although both transitions are expected to belong to the Gaussian universality class, the critical exponent associated with the one-particle gap closing is very different: For the AC one finds an exponent $z\nu\approx 1.472$ \cite{hu11} while an integer exponent $z\nu=1$ is expected for the DC \cite{haldane83a,haldane83b,botet83,schulz86,affleck87,kato94,yamamoto95,su16}. This results in a very different convergence behavior when one extrapolates the one-particle gap series for both spin-one chains using Pad\'{e} and DlogPad\'{e} extrapolation. We find that the extrapolation for the AC yields unsatisfactory results displaying a large uncertainty for the location of the critical point and for the value of the critical exponent. Only a biased Dlog-Pad\'{e} extrapolation shows more convincing values for the critical exponent approaching $z\nu \approx 1.5$. The situation is strongly different for the DC, although the calculated perturbative order is smaller. We find a very good convergence of Pad\'{e} and DlogPad\'{e} extrapolation for the one-particle gap. Our findings are in quantitative agreement with values of the literature. 

It is unclear what the reason is of the small discrepancies between the location of the quantum critical point for the DC when either studying the gap closing or physical properties of the ground state like entanglement measures is unclear. This question would deserve further investigations in the future.   

\acknowledgments
We thank F. Mila for fruitful discussions. DAR acknowledges support by the Max-Weber-Program in the Elite Network of Bavaria.

\bibliographystyle{apsrev4-1}


\end{document}